\documentclass{paper}
\usepackage{amsmath,amssymb,amsthm}
\usepackage{xspace,xifthen}
\usepackage{setspace}
\usepackage{apacite}
\usepackage{lineno}
\usepackage[hang,flushmargin]{footmisc}
\usepackage{etoolbox} 
\makeatletter
\patchcmd{\@footnotetext}{\@footnotemark}{\@footnotemark\onehalfspacing}{}{}
\makeatother
\usepackage{longtable}
\usepackage{booktabs}
\usepackage{array}

\newcommand{\cv}{Voting with Random Proposers\@\xspace}

\newcommand{\cw}{Condorcet winner\@\xspace}

\newcommand{\set}[1]{\ensuremath{\{#1\}}\xspace}

\newcommand{\al}[1][]{\ensuremath{x_{#1}}\xspace} 
\newcommand{\type}[1][]{\ensuremath{\theta_{#1}}\xspace} 
\newcommand{\cdf}[1][]
{
	\ifthenelse{\isempty{#1}}{
	\ensuremath{F\xspace}}{\ensuremath{F(#1)\xspace}}
} 

\newcommand{\pdf}[1][]{
	\ifthenelse{\isempty{#1}}{
	\ensuremath{f\xspace}}{\ensuremath{f(#1)\xspace}}
} 

\newcommand{\utility}[2]{\ensuremath{u_{#1}(#2)\xspace}} 
\newcommand{\med}{\ensuremath{\theta_{m}\xspace}} 
\newcommand{\mean}{\ensuremath{\mathbb{E}(\theta)\xspace}}
\newcommand{\mir}[1]{\ensuremath{c({#1})\xspace}} 

\newcommand{\q}[1][]{\ensuremath{q^{#1}\xspace}} 
\newcommand{\p}[1][]{\ensuremath{p^{#1}\xspace}} 
\newcommand{\w}[1][]{\ensuremath{w^{#1}\xspace}} 

\newcommand{\lbound}[1]{\ensuremath{\underline{#1}}\xspace}
\newcommand{\ubound}[1]{\ensuremath{\bar{#1}}\xspace}

\usepackage{pgfplots}
\usepackage{multirow}
\usepgfplotslibrary{fillbetween}
\DeclareMathOperator*{\argmax}{arg\,max}
\providecommand{\ie}{i.e.\@\xspace}
\providecommand{\href}[2]{#2}
\makeatletter
\@ifundefined{theorem}{\newtheorem{theorem}{Theorem}}{}
\@ifundefined{lemma}{\newtheorem{lemma}{Lemma}}{}
\@ifundefined{assumption}{\newtheorem{assumption}{Assumption}}{}
\makeatother
\begin{document}

%
%
\begin{titlepage}
\title{
	Voting with Random Proposers: Two Rounds May Suffice\footnote{We thank two anonymous referees and the advisory editor, as well as, Vincent Anesi, Pio Blieske, Jon X.\ Eguia, Sean Horan, Elias Kolleck, Markus Pivato, Clemens Puppe, Dimitrios Xefteris, Jan Zapal and workshop and seminar participants at the CEPS PolEcon Workshop in Paris 2024 and at the University of Zurich and ETH Zurich for valuable comments.}
}

\author{
	Hans Gersbach\thanks{KOF Swiss Economic Institute, ETH Zurich and CEPR, 
	\href{mailto:hgersbach@ethz.ch}
	{hgersbach@ethz.ch}.}
	\and
	Kremena Valkanova\thanks{Parity Technologies, 
	\href{mailto:kremena.valkanova@gmail.com}
	{kremena@parity.io}.}
	}

\date{This version: August 2026}

  \maketitle

\singlespacing

\begin{abstract}
	\noindent
This paper introduces the Voting with Random Proposers (VRP) procedure to address the challenges of agenda manipulation in voting. In each round of VRP, a randomly selected proposer suggests an alternative that is voted on against the previous round's winner. In a framework with single-peaked preferences, we show that the VRP procedure guarantees that the Condorcet winner is implemented in a few rounds, and in just two rounds under sufficiently symmetric preference distributions, or whenever the initial status quo or the first proposer's type is not extreme. Under the equilibrium proposal strategy, sincere and sophisticated voting induce the same winner in every round. The results have applications for committee decisions, legislative decision-making, and the organization of citizens’ assemblies and decentralized autonomous organizations. 
%
%
%
	\vspace{0.1in}
        \noindent\textbf{Conflicts of Interest:} None.\\
	\bigskip
\end{abstract}
\end{titlepage}
%
%

%
%
\doublespacing

\section{Introduction} \label{sec:introduction}
In various collective decision-making scenarios such as committee deliberations, legislative processes, and the governance of proof-of-stake blockchains, members of the corresponding bodies face the challenge of selecting a single alternative from a large set of options. A widely used approach in such contexts is to conduct successive rounds of pairwise majority voting \cite{Rasch2000}. For instance, legislative bodies often refine bills by sequentially voting on amendments, and negotiating parties in multi-agent resource allocation problems typically converge on a solution through incremental steps. However, with a large number of alternatives, it becomes infeasible to consider all options comprehensively. Moreover, the final outcome is highly sensitive to the subset of alternatives chosen for voting and the sequence in which they are presented. This creates a well-known vulnerability, whereby strategic monopoly agenda setters can manipulate the decision-making process to achieve outcomes aligned with their preferences, as extensively documented in the literature.\footnote{See the classic contributions by \citeA{McKelvey1979}, \citeA{Rubinstein1979}, \citeA{Bell1981}, \citeA{Schofield1983} and the more recent contributions that have extended and applied the potential and limit of agenda power to various settings, \citeA{Barbera2017}, \citeA{Nakamura2020} and \citeA{Nageeb2023}. We refer to \citeA{Horan2021} and \citeA{Rosenthal2022} for a recent account of the literature and to \citeA{Banks2002} for a review of the earlier literature.}

To address these challenges, we introduce an iterative majority voting procedure where, in each round, a \textit{randomly selected proposer} from the society suggests an alternative to be voted against the winner of the previous round. We refer to this method as ``Voting with Random Proposers" (VRP). The random selection of proposers is designed to decentralize and democratize the agenda-setting process, thus preventing any single agent from dominating the proposal sequence. Our analysis focuses on a simplified framework with a fixed number of voting rounds, one-dimensional alternatives, and a society of agents with symmetric single-peaked preferences with publicly known distribution. Additionally, only the final winning alternative is payoff relevant to the members of the society.

Our main goal is to analyze the strategic incentives created by the VRP procedure for both voters and proposers, while assessing its efficiency in selecting the socially optimal Condorcet winner. We show that an optimal proposal for a proposer in all rounds except the final one, given sophisticated voting, is the \cw, unless their peak lies at the tail of the preference distribution, in which case they propose their own peak. Conditional on this proposal strategy, sincere voting, \ie supporting whichever of the two standing alternatives a voter currently prefers, and sophisticated voting induce the same winner in every round. The procedure's performance therefore does not depend on voters' sophistication.

As a result, the probability of implementing the socially optimal alternative converges to one after only a few voting rounds.
For balanced distributions of the preferences, where the median and mean are sufficiently close, the Condorcet winner is obtained in only two voting rounds, independently of the status quo and the proposers' preferences. Even for arbitrary distributions of preferences in society, the \cw is selected in two rounds whenever the initial status quo or the first proposer's type is not extreme. The main intuition is that proposal-makers who are not in the final round face the following trade-off: if they propose a policy at or near their own peak, they risk that the final outcome will shift to the opposite side of the median. This risk can be avoided by proposing the median peak instead.

Our results demonstrate that current iterative voting procedures can be optimized by engaging randomly selected proposal-makers, thereby increasing the likelihood of reaching the \cw or at least come as close as possible. The VRP procedure has a wide range of potential applications, not only within standard collective decision-making bodies, such as committees and legislative bodies, but also in citizens’ assemblies that prepare proposals for city councils or parliaments. In this context, it may be optimal to select two randomly chosen citizen groups to develop proposals, which the assembly would then vote on, advancing the winning alternative to the next decision-making body. Moreover, this procedure could be effectively employed to decide on initiatives in direct democracies and can be easily adapted for decentralized autonomous organizations operating with distributed ledger technologies. 

\section{Relation to Literature}
Since the seminal result of \citeA{Moulin1980}, it is well known that in the case of single-peaked preferences, the Condorcet winner can be selected by asking agents to announce their peak, adding a number of fixed peaks, and using a suitable generalized median rule. \footnote{\citeA{Moulin1980} has shown that any strategy-proof social choice function on the domain of single-peaked preferences is a generalized median-rule that selects the median of the voters’ ballots and $n-1$ fixed ballots, where $n$ is the number of voters. However, not all generalized median rules select the Condorcet winner.} This result has been considerably extended by \citeA{Border1983}, \citeA{Barbera1994}, and \citeA{Klaus2020}. Unlike their mechanism design approach, our paper focuses on an alternative implementation problem: There is no central authority with commitment, i.e. a mechanism designer. The procedures can only involve proposal-making by agents and voting with equal proposal-making and agenda rights. Such procedures are sometimes called democratic mechanisms (see \citeA{Gersbach2008}) and also mirror common practices in parliamentary decision-making in representative democracies and referenda in direct democracies. In particular, we focus on procedures that present agents with two alternatives at a time.

Our paper also relates to the work of \citeA{Miller1977}, who demonstrates that a Condorcet winner is selected through sophisticated and potentially strategic voting, regardless of the agenda, as long as it includes the Condorcet winner. Similarly, \citeA{Austen1987} shows that strategic voting and sincere voting may be observationally equivalent since agenda-setters will propose options that can defeat the last winning proposal under sincere voting. We extend these insights by showing that the random selection of agenda-setters in an iterative majority voting process creates strong incentives for them to propose the Condorcet winner, even when they are self-interested and strategic. Under the equilibrium proposal strategy, sincere and sophisticated voting induce the same winner in every round, so implementation does not require voters to coordinate or to know each other's types.

Legislative bargaining models commonly feature random proposer selection and iterative majority voting on proposals, a framework originating from the seminal work of \citeA{Baron1989}. While much of the subsequent literature has focused on bargaining with an exogenous status quo\footnote{See \citeA{Eraslan2019} for a review.}, another line of research, notably \citeA{AnesiSeidmann2014}, explores dynamic bargaining models with an evolving status quo, where the status quo is determined by the outcome of the bargaining in the previous round.\footnote{This literature has been recently surveyed in \citeA{Eraslan2022}.} Our model features several important deviations from these frameworks, including a finite number of voting rounds, a one-dimensional policy space\footnote{In this respect, our paper relates to the literature on spatial bargaining, including \citeA{Baron1991}, \citeA{Banks2000}, \citeA{Kalandrakis2016}, and \citeA{Zapal2016}.}, and a single payoff-relevant outcome. Within this context, our model can be classified as a spatial legislative bargaining model featuring a finite horizon and an evolving status quo, where the winning  alternative of each round becomes the status quo for the next round but is only implemented at the final period, akin to amendment procedures in legislatures \cite{Eraslan2022}. We obtain moderation due to the threat of future polarization: If I know that my ideological opponent will make a proposal tomorrow, I have an incentive to constrain her through my proposal today. In the presence of a decisive median voter, the way to impose such a constraint is to propose a policy that appeals to the median.

Our paper also relates to the literature on strategic polarization. \citeA{Kalai2001} demonstrate that strategic behavior within aggregation games tends to foster polarization. In comparison, our work illustrates that well-crafted institutional mechanisms, such as the Voting with Random Proposers procedure, can promote moderation and facilitate convergence toward compromise, more exactly the \cw. The intuition behind this contrast is the following. Polarization arises when each player directly influences the aggregate outcome. VRP, however, counters this by removing direct aggregation and instead using randomized sequential proposals with majority voting. In another work, \citeA{Eraslan2020} show that equilibrium outcomes evolve gradually through repeated majority voting, but the process may take many rounds. In comparison, our investigations demonstrate that randomizing proposers aligns incentives so that the \cw emerges almost immediately, often within just two rounds. Thus, while the former relies on gradual stabilization through voting dynamics, the latter achieves rapid convergence through institutional design of the proposal stage.

Our research on voting with random proposers over a large set of alternatives also relates to the literature on algorithmically determined and iterative voting procedures with a random component in various policy spaces, as developed by \citeA{Airiau2009}, \citeA{Goel2012} and \citeA{Garg2019}, among others. We add to this literature by studying a simple procedure to implement the \cw. 

Finally, when designing iterative voting procedures with endogenous proposal-making, it is crucial to identify alternatives in each round that are closer to the Condorcet winner.\footnote{We use reaching the \cw as the objective of designing voting procedures.} In this sense, our work contributes to the literature by \citeA{Callander2011}, \citeA{Riboni2010}, and \citeA{Barbera2022}, which examines how to identify desirable policies to put to a vote. We demonstrate that the sequential random selection of proposers plays a pivotal role in guiding proposers toward making socially superior proposals.
%
%

%
%
\section{The Model}
\label{sec-model}
Consider a society that consists of a continuum of agents with mass one, each of them privately informed about their type $\theta\in [0,1]$. Agents' types are drawn from a publicly known cumulative distribution function $F$. We assume throughout that $\cdf[]$ is absolutely continuous on $[0,1]$ with density $\pdf[]$, and that its median $\med$ is uniquely determined, i.e.\ $\cdf[\med]=\tfrac12$ and $\cdf[]$ is strictly
increasing in a neighborhood of $\med$. These conditions guarantee that the median peak is well defined and that majorities in a neighborhood of $\med$ are decisive. 

The utility of an agent of type $\type$ from an alternative $\al\in [0,1]$ is given by $\utility{\type}{\al}=-(\al-\type)^2$, therefore it has a unique maximum at $\type$ and is symmetric around $\type$. Since preferences are single-peaked, the median peak $\med$ is the Condorcet winner \cite{Black1948}.

We define a function $c:[0,1]\rightarrow [0,1]$, $x\mapsto\min\set{\max\set{2\med-x,0},1}$. Thus, $c(x)$ is the farthest positioned alternative in $[0,1]$ that is at least as preferred as $x$ by a majority of society.
Note in particular that whenever $x\in(\max\{0,2\med-1\},\min\{2\med,1\})$, we have $F\left(\frac{x+c(x)}{2}\right)=\cdf[\med]=\frac{1}{2}$. 

Furthermore, we define lower and upper thresholds as the boundaries of the regions in which a proposer strictly prefers the lottery induced by their own peak to the certain outcome $\med$:
\begin{align*}
\lbound{\type}
&=\sup\left(
\{0\}\cup
\left\{
\type\in[0,1]:
\type\leq 2\med-1,\ 
\int_{\type}^{1}\utility{\type}{v}\pdf[v]\,dv
>\utility{\type}{\med}
\right\}
\right),\\
\ubound{\type}
&=\inf\left(
\{1\}\cup
\left\{
\type\in[0,1]:
\type\geq 2\med,\ 
\int_{0}^{\type}\utility{\type}{v}\pdf[v]\,dv
>\utility{\type}{\med}
\right\}
\right).
\end{align*}
Intuitively,  a type below $2\med-1$, the lower-tail comparison is between the certain utility $\utility{\type}{\med}$ and $\int_{\type}^{1}\utility{\type}{v}\pdf[v]\,dv$, the expected utility when the alternative is drawn from $\cdf$ but truncated below at the proposer's peak. The upper-tail comparison is symmetric. Types strictly below $\lbound{\type}$, or strictly above $\ubound{\type}$, strictly prefer the lottery induced by their own peak; at a threshold, and whenever the corresponding strict-preference region is empty, we select $\med$. When $\med<\tfrac12$, no type satisfies $\type\leq2\med-1$; at $\med=\tfrac12$ the only admissible point is $\type=0$. Thus $\lbound{\type}=0$ whenever $\med\leq\tfrac12$; symmetrically, when $\med>\tfrac12$, $\ubound{\type}=1$.

We observe that the remaining threshold also collapses when the most extreme type on the relevant side weakly prefers the median peak to the lottery induced by its own peak: $\lbound{\type}=0$ when $\med>\tfrac12$ and
\[
\int_0^1\utility{0}{v}\pdf[v]\,dv\leq\utility{0}{\med},
\]
and, symmetrically, $\ubound{\type}=1$ when $\med\leq\tfrac12$ and
\[
\int_0^1\utility{1}{v}\pdf[v]\,dv\leq\utility{1}{\med}.
\]

We require $\cdf[]$ to satisfy the following additional assumptions. For $h\in[0,H]$, where $H:=\min\set{\med,1-\med}$,
\[
q(h):=1-\cdf[\med+h]-\cdf[\med-h],
\qquad
s(h):=1-\cdf[\med+h]+\cdf[\med-h].
\]
Both are built from the two tail masses $1-\cdf[\med+h]$ (above $\med+h$) and
$\cdf[\med-h]$ (below $\med-h$): $q(h)$ is their \emph{difference}, a signed measure of
tail imbalance that is positive when the upper tail is the heavier of the two at 
$h$, and $s(h)\ge 0$ is their \emph{sum}, the combined mass in both tails. Our first restriction signs the tail imbalance in the aggregate.
 
\begin{assumption}
\label{assumption:F}
The distribution of types satisfies
\[
\int_{0}^{\med} q(h)\,dh\ \ge\ 0 \quad\text{if } \med\le\tfrac12,
\qquad\text{or}\qquad
\int_{0}^{1-\med} q(h)\,dh\ \le\ 0 \quad\text{if } \med>\tfrac12 .
\]
\end{assumption}
 
\noindent
The inequality compares the two tails in the aggregate: for $\med\le\tfrac12$ it holds when
the upper tail is at least as heavy as the lower one, ruling
out a disproportionate mass near the lower boundary. As we show in
Lemma~\ref{lem-mean-median}, this is exactly what places the mean on the appropriate side
of the median.\footnote{A more classical sufficient condition is to require $q(h)$ to have
constant sign for all $h$ rather than only after integration. This is equation~(1.2) of
\citeA{vanZwet1979}. Our Assumption~\ref{assumption:F} (its integrated form) is
equation~(1.1). We use the weaker integral condition because it is all the results require.}


\begin{lemma}
\label{lem-mean-median}
Under Assumption~\ref{assumption:F},
$\mean-\med \geq \int_{2\med}^{1}\!\bigl[1-F(v)\bigr]\,dv\geq 0,$ if $\med \leq \tfrac{1}{2}$,  and 
$\mean-\med \leq -\int_{0}^{2\med-1}F(v)\,dv\leq 0$ if $\med > \tfrac{1}{2}$.
\end{lemma}
\begin{proof}
See Appendix~\ref{proof-mean-median}.
\end{proof}
Assumption~\ref{assumption:F} constrains only the \emph{direction} of the imbalance, that
is, the sign of its integral. In addition, we impose a
bound on the \emph{magnitude} of the accumulated imbalance relative to the accumulated
dispersion. 
 
\begin{assumption}
\label{assumption:F3}
For all $h\in[0,H]$,
\[
\int_{0}^{h} r\,s(r)\,dr\ \ge\ -\big(\med-\lbound{\type}\big)\!\int_{0}^{h} q(r)\,dr,
\qquad\text{and}\qquad
\int_{0}^{h} r\,s(r)\,dr\ \ge\ \big(\ubound{\type}-\med\big)\!\int_{0}^{h} q(r)\,dr.
\]
\end{assumption}
 
\noindent
Assumption~\ref{assumption:F} signs the single quantity
$\int_0^h q$ at the outer scale $h=H$ and so fixes the direction of the skewness,
Assumption~\ref{assumption:F3} governs the entire profile $\{\int_0^h q\}_{h\le H}$ and
bounds its size relative to the dispersion. The two are complementary, and neither implies
the other. Both hold with room to spare when $\cdf$ is close to symmetric about $\med$ (if
$\cdf$ is exactly symmetric then $\int_0^h q\equiv 0\le\int_0^h r\,s$, so both are
slack) and Assumption~\ref{assumption:F3} can fail only for distributions whose tails are
strongly imbalanced relative to their spread.

\subsubsection*{Voting with Random Proposers}

The society uses a multi-round voting procedure called \cv to implement an alternative from the interval $[0,1]$ in a publicly known fixed number of voting rounds~$T\in\set{2,3,\dots}$. Suppose that initially, there is a status quo alternative $\q[1]\in[0,1]$. In round $t=1$, a proposer of type $\type[s^1]$ is randomly drawn from the distribution~$F$ and makes a proposal $\p[1]\in[0,1]$. Then, a majority-vote between \p[1] and \q[1] takes place, where ties are resolved in favor of the proposal. All voters in the society take part in each round of voting. The winner of the first voting round~$\w[1]$ becomes the status quo in the next round, \ie $\w[1]=\q[2]$. In round $t=2$, a new proposer is randomly drawn and the process is repeated. The procedure ends at round~$T$ and the winner of the last round \w[T] is implemented. Proposers and voters are assumed to be strategic, \ie they would choose a proposal or vote in favour of alternatives in order to maximize the proximity of the final winner to their type.

\subsubsection*{Equilibrium concept}

The VRP procedure defines a finite-horizon dynamic Bayesian game. In principle, the appropriate solution concept is Perfect Bayesian Equilibrium (PBE). However, under the assumptions of the model, the beliefs are simple and remain history independent and the equilibrium analysis reduces to finding Subgame Perfect Nash Equilibria (SPNE) for the following reasons.
Although types are privately known and the game is formally one of incomplete information, no payoff-relevant variable is ever uncertain. A player's payoff depends only on the implemented alternative, which depends on the type profile only through the distribution~$\cdf[]$. Since~$\cdf[]$ is common knowledge and we adopt the standard convention that the empirical distribution of types in the nonatomic population coincides with~$\cdf[]$, so
that the realized median equals~$\med$, the entire payoff-relevant environment is common knowledge. A proposal may reveal the proposer's own type, but that type is payoff-irrelevant to every other agent, is uninformative about any other type by independence, and, having measure zero, leaves~$\cdf[]$ unchanged. The proposal-signaling channel of \citeA{Banks1990,Banks1993} is therefore absent. Likewise, no round outcome can revise beliefs about~$\med,$ so the voter-signaling channel of \citeA{Piketty2000} and \citeA{Razin2003} is absent as well. Beliefs about payoff-relevant variables thus coincide with the prior at every information set, and PBE reduces to SPNE.\footnote{With finitely many voters the realized median is no longer common knowledge, and proposals and votes become informative about it. The analysis would then require Perfect Bayesian Equilibrium with explicit beliefs and would face the multiplicity issues documented in this literature.}

Because an individual voter is non-pivotal in a nonatomic electorate, SPNE alone does not discipline individual ballots. We therefore impose a \emph{pivotal-voting refinement}: at every binary voting information set, each voter chooses the alternative that gives her the higher continuation utility under the assumption that her ballot determines the winner. We call this \emph{sophisticated voting}. This refinement is the nonatomic analogue of the pivotal-voter logic used in finite voting games \cite{Baron1993}.

An SPNE satisfying the pivotal-voting refinement is called an equilibrium in the remainder of the paper. This refinement evaluates each ballot in the decision problem conditional on pivotality; no unconditional dominance claim is made for an individual voter in the nonatomic game. A voter votes \emph{sincerely} in such an equilibrium if she supports whichever of the two standing alternatives she currently prefers, \ie a voter of type $\type$ votes for $\p[t]$ whenever $|\type-\p[t]|<|\type-\q[t]|$ and for $\q[t]$ whenever $|\type-\p[t]|>|\type-\q[t]|$. A set of voters is \emph{decisive} in a round if the alternative it supports wins regardless of the remaining votes: it must have mass at least $\tfrac12$ when it supports the proposal and mass strictly greater than $\tfrac12$ when it supports the status quo.

%
%
%
%
\section{Equilibrium Proposals and Implementation}

In this section, we study the optimal proposal of the proposers, depending on their type, the distribution of preferences, and the status quo, as well as the outcome of the VRP procedure. As we show in the theorem below, the equilibrium strategy for all proposers in rounds $t \in \set{1,2,\dots, T-1}$ is identical under sophisticated voting, and they either propose their own type or the Condorcet winner. In the last round of voting, the proposer proposes either their own type or the farthest positioned alternative that wins in a pairwise vote against the status quo. 
\begin{theorem}
\label{prop-random-cw}
Let Assumptions~\ref{assumption:F} and~\ref{assumption:F3} hold. Given sophisticated voting, the following rule selects an optimal proposal for every proposer. In rounds $t\in\set{1,2,\dots,T-1}$
\begin{align*}
\p[t]=
\begin{cases}
\type[s^{t}]& \text{ if }\max\set{\q[t], \type[s^{t}]}< \lbound{\type} \text{ or }  \min\set{\q[t],\type[s^{t}]}>\ubound{\type},\\
\med & \text{else.}
\end{cases} \text{ and } & \p[T]=\begin{cases}
\min\set{\type[s^T],\mir{\q[T]}}& \text{if } \q[T]\leq \med,\\
\max\set{\type[s^T],\mir{\q[T]}}& \text{if } \q[T]> \med.
\end{cases}
\end{align*}
Under this proposal strategy, sincere and sophisticated voting induce the same winner in every round.
\end{theorem}

\begin{proof}
    See Appendix~\ref{proof-random-cw}.
\end{proof}

Proposers face a trade-off between proposing their own type to maximize their utility and proposing the \cw to insure themselves against unfavorable final winners. The closer a proposer’s type is to the \cw, the stronger the incentive to propose it. The share of such voters increases with the symmetry of the distributions of the types. However, in sufficiently skewed distributions, the \cw is positioned closer to one extreme, leading proposers with preference peaks at the tail of the distribution to favor their own type. This is the case, because the potential benefit from having all subsequent proposers positioned at the tail outweighs the insurance gained by proposing the \cw, which is nearly as disliked as the extreme opposite alternatives.\footnote{The symmetry of the utility and its dependence on $x$ and $\theta$ only through the distance $|x-\theta|$ are what make the median peak the proposers' insurance-optimal proposal and pin down the threshold type $\lbound{\type}$. The curvature of the loss reshapes these
thresholds and sets the strength of the distributional condition, but not the
qualitative result: a flatter loss tightens the condition and must be offset by a
more symmetric $\cdf[]$, until in the linear limit only $\mean=\med$ remains.} The optimal strategy of the proposers is illustrated in Figure~\ref{fig-types}, depending on the distribution of the types. 

\begin{figure}[!h]
\begin{center}
\begin{tabular}{@{}c @{}c@{}c}
	$\theta\sim \operatorname{Beta}(4,2)$ & $\theta\sim \operatorname{Beta}(20,2)$ &$\theta\sim \operatorname{Beta}(0.3,0.2)$ \\
	\begin{tikzpicture}[scale=1]
    \begin{axis}[axis x line*=bottom, axis y line*=left, xlabel={$\theta$}, ylabel={$f(\theta)$},width=5.9cm,xtick={0,0.2,0.4,0.6,0.8,1},
xticklabels={0,0.2,0.4,0.6,0.8,1}]
        \addplot[domain = 0:1, name path=f] {20*(1 - x)*x^3};
        \addplot[domain = 0:1,draw=none,name path=B] {0};     
        \addplot[fill=CW] fill between[of=f and B, soft clip={domain=0:1}];
        \draw [dashed, black]  (0.68619,0) -- (0.68619,4.5);
    \end{axis}
	\end{tikzpicture}  & 
	\begin{tikzpicture}[scale=1]
    \begin{axis}[axis x line*=bottom, axis y line*=left,xlabel={$\theta$},width=5.9cm,xtick={0,0.2,0.4,0.6,0.8,1},
xticklabels={0,0.2,0.4,0.6,0.8,1}]
        \addplot[domain = 0:1, name path=f] {420*(1 - x)*x^19};
        \addplot[domain = 0:1, draw=none,name path=B] {0};     
        \addplot[CW] fill between[of=f and B, soft clip={domain=0.77295:1}];
        \addplot[OWN] fill between[of=f and B, soft clip={domain=0:0.77295}];
        \draw [dashed, black]  (0.921356,0) -- (0.921356,10);
    \end{axis} 
	\end{tikzpicture} &
		\begin{tikzpicture}[scale=1]
    \begin{axis}[axis x line*=bottom, axis y line*=left, xlabel={$\theta$},width=5.9cm,xtick={0,0.2,0.4,0.6,0.8,1},
xticklabels={0,0.2,0.4,0.6,0.8,1}]
        \addplot[domain = -0.01:1.1, , name path=f] {0.12905754687023/((1 - x)^0.8*x^0.7)};
        \addplot[domain = 0:1, draw=none,name path=B] {0};     
        \addplot[fill=CW] fill between[of=f and B, soft clip={domain=0.207543:1}];
        \addplot[fill=OWN] fill between[of=f and B, soft clip={domain=0:0.207543}];
                \draw [dashed, black]  (0.760865,0) -- (0.760865,4.5);
    \end{axis}
	\end{tikzpicture}\\
 	\begin{tikzpicture}[scale=1]
   \begin{axis}
        [axis x line*=bottom, axis y line*=left,
        xlabel={$T$},
        ylabel={Prob($w^T=\med$)},
        xmin=1.7, xmax=5.1,
        ymin=0.7, ymax=1,,width=5.9cm,
        every axis plot post/.style={mark options={black}}
        ]
            \addplot+[ycomb, domain=2:5, samples at={2,3,4,5}, black]{1};
    \end{axis}
	\end{tikzpicture}  & 
	\begin{tikzpicture}[scale=1]
    \begin{axis}
        [axis x line*=bottom, axis y line*=left,
        xlabel={$T$},
        xmin=1.7, xmax=5.1,
        ymin=0.7, ymax=1,width=5.9cm,
        every axis plot post/.style={mark options={black}}
        ]
            \addplot+[ycomb, domain=2:5, samples at={2,3,4,5}, black]{1 - 0.03210800826907127^(x - 1)};
    \end{axis}
	\end{tikzpicture} &
		\begin{tikzpicture}[scale=1]
    \begin{axis}
        [axis x line*=bottom, axis y line*=left,
        xlabel={$T$},
        xmin=1.7, xmax=5.1,
        ymin=0.7, ymax=1,width=5.9cm,
        every axis plot post/.style={mark options={black}}
        ]
            \addplot+[ycomb, domain=2:5, samples at={2,3,4,5}, black]{1 - 0.279950554697268^(x - 1)};
    \end{axis}
	\end{tikzpicture}
\end{tabular}
\caption{The first row depicts the selected optimal proposals in rounds $t\in\set{1,2,\dots,T-1}$ for different Beta distributions of types. Types in the green regions propose the \cw, and those in the red regions propose their own type. The dashed line marks the position of the \cw. The regions are drawn for a status quo lying in the same extreme region as the proposer; whenever the status quo lies in $[\lbound{\type},\ubound{\type}]$, every type proposes the \cw. The second row shows the probability of implementing the \cw for $T\in\{2,\dots,5\}$ for each distribution.}
\label{fig-types}
\end{center}
\end{figure}
This proposal strategy is supported by sophisticated voting. As shown in Lemma~\ref{lem3} in the Appendix, sincere and sophisticated voting induce the same winner in every round under this proposal strategy, so the result does not depend on voters' sophistication.

It follows from the proposers' equilibrium strategy that, for highly skewed distributions of types and a status quo positioned at the tail of the distribution, the final winner may itself lie at the tail. This possibility becomes progressively less likely as the number of rounds grows, because
sustaining a tail outcome requires every subsequent proposer to be drawn from the tail. Consequently, the probability of implementing the \cw approaches one within a few rounds, as illustrated in Figure~\ref{fig-types}. The following theorem makes this precise: part~(i) gives the exact rate at which this probability converges to one, and part~(ii) identifies conditions under which VRP always the Condorcet winner with certainty. Part~(ii) relies on the following restriction on the distribution of types.

\begin{assumption}
\label{assumption:F2}
The distribution of types $\cdf[]$ satisfies
\begin{equation*}
\text{Var}(\type)\geq
\begin{cases}
\med^2-\mean^2 & \text{if }\med\geq 0.5,\\[2pt]
(1-\med)^2-(1-\mean)^2 & \text{if }\med<0.5.
\end{cases}
\end{equation*}
\end{assumption}

The condition holds whenever the median and the mean are sufficiently close, that is, whenever the distribution is not highly asymmetric relative to its dispersion.

\begin{theorem}
\label{prop-2rounds}
Let Assumptions~\ref{assumption:F} and~\ref{assumption:F3} hold, and consider the proposal strategy of Theorem~\ref{prop-random-cw}.
\begin{enumerate}
\item[(i)] If $\q[1]\notin[\lbound{\type},\ubound{\type}]$,
\begin{align*}
\text{Prob}(\w[T]=\med)=
\begin{cases}
1-\cdf[\lbound{\type}]^{\,T-1} & \text{if } \med\geq 0.5,\\[2pt]
1-\bigl(1-\cdf[\ubound{\type}]\bigr)^{T-1} & \text{if } \med<0.5.
\end{cases}
\end{align*}
For any $\q[1]\in[\lbound{\type},\ubound{\type}]$, we have $\text{Prob}(\w[T]=\med)=1$.
\item[(ii)] If $\q[1]\in[\lbound{\type},\ubound{\type}]$ or $\type[s^{1}]\in[\lbound{\type},\ubound{\type}]$, then $\w[T]=\med$ for every $T\geq 2$. If Assumption~\ref{assumption:F2} holds in addition, then $\w[T]=\med$ for every $\q[1]$, $\type[s^{1}]$ and $T\geq 2$.
\end{enumerate}
\end{theorem}


\begin{proof}
    See Appendix~\ref{proof-2rounds}.
\end{proof}
Part~(i) shows that the probability of selecting the \cw rises geometrically in
the number of rounds. When $\med>0.5$, it equals
$1-F(\lbound{\type})^{T-1}\geq 1-F(\med)^{T-1}$, since $\lbound{\type}\leq\med$,
with the symmetric bound applying when $\med<0.5$. Taking $1-F(\med)^{T-1}$ as a
distribution-free lower bound, the \cw is implemented with probability exceeding
$96.8\%$ within six voting rounds, regardless of the specific distribution of
types.

Part~(ii) identifies conditions under which the \cw is proposed in the first round, wins the first vote and is never displaced, so that it is implemented for any horizon, in particular for $T=2$. This is the case whenever the first-round status quo or the first proposer's type is not extreme, i.e.\ lies in $[\lbound{\type},\ubound{\type}]$. Under Assumption~\ref{assumption:F2} the threshold interval is the whole policy space, so the condition holds for every status quo and every proposer type. Intuitively, under Assumption~\ref{assumption:F2} all types prefer to propose the \cw rather than their own peak, since otherwise the next proposer might implement a policy on the opposite side of the distribution of peaks. Thus, the threat of a polarized outcome tomorrow induces moderation in today's policies.

Assumption~\ref{assumption:F2} is the incentive constraint of
the most extreme type. Suppose $\med>0.5$, so that $\type[s]=0$ is the type
furthest from the \cw. Proposing $\med$ delivers the certain loss $\med^{2}$,
whereas proposing its own peak leaves the final winner distributed as $\cdf[]$
on $[0,1]$, with expected loss $\text{Var}(\type)+\mean^{2}$. The extreme type
faces a trade off between location and risk: the lottery is centered at $\mean\leq\med$ and so
lies nearer its peak, but it is dispersed. It proposes the \cw precisely when the
dispersion it would bear outweighs this location advantage,
$\text{Var}(\type)\geq\med^{2}-\mean^{2}$, which is Assumption~\ref{assumption:F2}.
Assumption~\ref{assumption:F} makes the lottery attractive in location, since it
implies $\mean\leq\med$, whereas Assumption~\ref{assumption:F2} requires the dispersion
to be large enough that the extreme type nonetheless prefers the certain outcome.
Equivalently, Assumption~\ref{assumption:F2} holds if and only if
$\lbound{\type}=0$ (respectively $\ubound{\type}=1$ when $\med<0.5$), \ie
when even the most extreme type weakly prefers the \cw, so the green region in
Figure~\ref{fig-types} covers the whole support.\footnote{The two distributional conditions are related but not nested. Assumption~\ref{assumption:F2} compares the most extreme type with the lottery generated by the full distribution $\cdf[]$, whereas Assumption~\ref{assumption:F3} compares the threshold types with lotteries generated by distributions censored to symmetric windows about the median. The underlying lotteries differ, so neither assumption is an instance of the other.} Although Assumptions~\ref{assumption:F2} and~\ref{assumption:F3} are not nested, Assumption~\ref{assumption:F2} makes $[\lbound{\type},\ubound{\type}]=[0,1]$. It therefore implies that the realisation-level sufficient condition in Theorem~\ref{prop-2rounds}(ii) holds for every status quo and proposer type. When Assumption~\ref{assumption:F2} fails, that condition may still deliver two-round implementation for particular realisations.

Together the two assumptions bound the nonparametric skewness
$S:=(\mean-\med)/\text{Std}(\type)$ of $F$. In particular, for $\med>0.5$:
\begin{equation*}
\frac{1}{\text{Std}(\type)}\int_{0}^{2\med-1}\!F(v)\,dv
\;\leq\; -S \;\leq\; \frac{\text{Std}(\type)}{\med+\mean},
\end{equation*}
where the lower and upper bounds follow from Assumptions~\ref{assumption:F} (see Lemma~\ref{lem-mean-median}) and~\ref{assumption:F2}, respectively. Analogous bounds hold for $\med<0.5$. The upper bound reflects the quadratic loss: the extreme type weighs
a squared distance against a second moment, so the admissible skewness rises with
the standard deviation. It is therefore not a universal cap on $S$: a tightly
concentrated $F$ must be nearly symmetric, whereas a dispersed one may be
substantially skewed.

\section{Discussion and Outlook}
We demonstrate that incorporating random proposers into an iterative majority voting process efficiently implements the socially optimal alternative within a few rounds. Furthermore, if individual utility-maximizing alternatives are sufficiently symmetrically distributed, or if the initial status quo or the first proposer's type is not extreme, the VRP procedure selects the \cw in only two rounds.

We use a simple framework with three key assumptions. First, the distribution of peaks of the individual preferences is public knowledge. 
Second, there is a continuum of voters, which ensures that no single voter can be selected as proposer more than once and that proposals carry no private information, thereby eliminating signaling incentives.
Third, our analysis assumes single-peaked preferences. 

Our results also suggest that the central moderation mechanism may extend, at least approximately, when the distribution of preference peaks is not publicly known. In such an environment, the relevant benchmark would no longer be the known median $\theta_m$, but the median of the posterior distribution over preference peaks. A proposer who is uncertain about the location of the Condorcet winner may still have an  incentive to propose an alternative close to this posterior
median in order to insure herself against an unfavorable proposal in a later round. The main difference is that proposals and voting outcomes may now convey information about the distribution of preferences, so that the relevant posterior and its median can evolve across rounds. The qualitative force toward
moderation may therefore survive, while exact implementation of the Condorcet winner may require additional rounds and depend on the informativeness of the observed history. Formally characterizing this interaction between learning, proposal incentives, and convergence under VRP is a natural direction for future research.

A related extension concerns finite electorates. With finitely many voters, the realized median is uncertain, individual voters may be pivotal, and a proposal can reveal information about both the proposer and the composition of the electorate. Consequently, the exact continuum results need not hold: proposal-making may involve signaling \cite{Banks1990,Banks1993}, voting may reflect both pivotal and informational considerations \cite{Piketty2000,Razin2003}, and equilibrium outcomes may be history dependent. Nevertheless, when the electorate is large and the realized median is sufficiently concentrated around the population median, the continuum model suggests that VRP may retain an approximate implementation property, except in histories in which the relevant majority comparisons are close. Future work could establish finite-population bounds on the probability of implementing a
Condorcet winner, study how rapidly these bounds approach the continuum
benchmark, and determine whether alternative disclosure or proposer-selection rules can limit the additional signaling incentives.

Finally, our analysis assumes single-peaked preferences. It is well known that a \cw might not exist when preferences violate this assumption and pairwise voting procedures might exhibit cycles. \cv offers a probabilistic approach to resolving such cycles. Future research should build on the work of \citeA{Airiau2009}, who show that the proposal dynamic resembles a Markov chain, and explore the convergence and strategic incentives in more general preference domains.


%
%

\newpage

\bibliographystyle{apacite}
\bibliography{vrp}
\newpage

\clearpage

\renewcommand{\arraystretch}{1.15}
\section*{Notation}

\begin{longtable}{@{}p{3cm}p{11cm}@{}}
\toprule
\textbf{Variable} & \textbf{Explanation} \\
\midrule
\endfirsthead
\toprule
\textbf{Variable} & \textbf{Explanation} \\
\midrule
\endhead
\bottomrule
\endfoot

$[0,1]$ & Policy space \\
$\theta$ & Type of an agent in $[0,1]$, representing his/her peak of preferences \\
$F$ & Cumulative distribution function of types \\
$f$ & Density function of types \\
$\med$ & Median of the type distribution \\
$\mathbb{E}(\theta)$ & Mean of the type distribution \\
$\mathrm{Var}(\theta)$ & Variance of the type distribution \\
$\operatorname{Beta}(\alpha,\beta)$ & Beta distribution with parameters $\alpha$ and $\beta$ \\
$u_\theta(x)$ & Utility of type $\theta$ for policy $x$ \\
$c(x)$ & Reflection of $x$ around $\med$, truncated to $[0,1]$ \\
$\underline{\theta}$ & Lower threshold type \\
$\bar{\theta}$ & Upper threshold type \\
$T$ & Fixed number of voting rounds \\
$q^t$ & Status quo in round $t$; $q^{t+1}=w^t$ \\
$p^t$ & Proposal made in round $t$ by the randomly drawn proposer \\
$w^t$ & Winner of the pairwise majority vote in round $t$ \\
$\theta_{s^t}$ & Type of the proposer drawn in round $t$ \\
$q(h)$ & Difference between the upper- and lower-tail masses at distance $h$ from $\med$ \\
$s(h)$ & Sum of the two tail masses at distance $h$ from $\med$ \\
$H$ & Maximum symmetric window width, $H=\min\{\med,1-\med\}$ \\
$X_h$ & Round-$T$ proposer type censored to $[\med-h,\med+h]$ \\
$S$ & Nonparametric skewness measure, $(\mean-\med)/\mathrm{Std}(\type)$ \\

\end{longtable}

\clearpage

\appendix
%
%
%

\section{Appendix: Proof of Lemma~\ref{lem-mean-median}}
\label{proof-mean-median}

Since $\med$ is a constant, we express the difference between the mean and the median using the tail-integral representation:
\begin{align*}
\mean-\med=\mathbb{E}(\theta-\med)
&=\int_0^{\infty}\Pr(\theta-\med>h)\,dh-\int_{0}^{\infty}\Pr(\theta-\med<-h)\,dh \\
&=\int_0^{1-\med}\!\bigl[1-F(\med+h)\bigr]\,dh
-\int_0^{\med}\! F(\med-h)\,dh,
\end{align*}
where the limits tighten because $1-\cdf(\med+h)=0$ for $h>1-\med$ and
$\cdf(\med-h)=0$ for $h>\med$, given $\type\in[0,1]$.

Consider first a distribution with $\med\leq\tfrac12$, so that $\med\leq 1-\med$. We split the upward integral at $h=\med$ and group its lower part with the downward integral:
\[
\mean-\med
=\int_0^{\med}\!\bigl[1-F(\med+h)-F(\med-h)\bigr]\,dh
+\int_{\med}^{1-\med}\!\bigl[1-F(\med+h)\bigr]\,dh .
\]
Since the first integral is nonnegative by Assumption~\ref{assumption:F}, we have 
\[
\mean-\med \geq \int_{\med}^{1-\med}\!\bigl[1-F(\med+h)\bigr]\,dh = \int_{2\med}^{1}\!\bigl[1-F(v)\bigr]\,dv\geq 0,
\]
where the equality results from substituting $v=\med +h$ and is the final result.
The case $\med>\tfrac12$ is handled the same way, with the roles of the two tails reversed. Since
\[
\mean-\med
=\int_0^{1-\med}\!\bigl[1-F(\med+h)-F(\med-h)\bigr]\,dh
-\int_{1-\med}^{\med}\! F(\med-h)\,dh,
\]
and the first integral is nonpositive by Assumption~\ref{assumption:F}, 
\[
\mean-\med \leq -\int_{1-\med}^{\med}\! F(\med-h)\,dh = -\int_{0}^{2\med-1}F(v)\,dv\leq 0,
\]
where the equality results from substituting $v=\med -h$ and is the final result.

\section{Appendix: Proof of Theorem~\ref{prop-random-cw}}	
\label{proof-random-cw}
Recall that $\mir{x}$ denotes the farthest positioned alternative in $[0,1]$ that is at least as preferred as $x$.
It is given by the symmetric point of $x$ with respect to the \cw, \med, \ie, $\mir{\al}:=\min\set{\max\set{2\med-\al,0},1}$.
We solve the sequential game through backward induction starting at round $T$.\\
\textbf{Optimal proposal in round $T$}\\
In the last round~$T$ the winner is implemented immediately, so that every voter's
continuation utility from an alternative coincides with her current utility from it.
Sincere and sophisticated voting therefore coincide for \emph{every} voter in
round~$T$, irrespective of the alternatives on the ballot, and each voter supports
whichever of $\p[T]$ and $\q[T]$ is closer to her peak. Since preferences are
single-peaked and the round is a pairwise majority vote, the winning alternative
$w(\p[T],\q[T])$ for $\q[T]\leq \med$ is given by
\begin{equation}
w(\p[T],\q[T])=\begin{cases}
\p[T] &\text{ if } \p[T]\in[\q[T],\mir{\q[T]}],\\
\q[T] &\text{ if } \p[T]\in[0,\q[T])\cup(\mir{\q[T]},1].
\end{cases}
\end{equation}
The peak of the preference of the proposer at round $T$ is given by $\type[s^T]$. Consider the optimization problem of the proposer in round $T$:
	\begin{equation*}
	\argmax_{\p[T]\in[\q[T],\mir{\q[T]}]} - (\p[T]-\type[s^T])^2.
	\end{equation*}
Note that the proposer is not constrained about the position of the proposal in the interval $[0,1]$ by the voting procedure, but he/she would anticipate that proposing an alternative in $[0,\q[T])$ or $(\mir{\q[T]},1]$ would lose the pairwise vote and cannot be utility-improving.
Thus, the optimal proposal and final winner are given by
\begin{align*}
\p[T]\in\begin{cases}
[0,\q[T]]& \text{if } \type[s^T]\in[0, \q[T]],\\
\set{\type[s^T]}	& \text{if } \type[s^T]\in[\q[T],\mir{\q[T]}],\\
\set{\mir{\q[T]}	}	&\text{else}.
\end{cases}\qquad\qquad&w(\p[T],\q[T])=\begin{cases}
\q[T]& \text{if } \type[s^T]\in[0, \q[T]],\\
\type[s^T]	& \text{if } \type[s^T]\in[\q[T],\mir{\q[T]}],\\
\mir{\q[T]} &\text{else}.
\end{cases}
\end{align*}

Whenever $\type[s^T]\in[0,\q[T])$, every proposal in $[0,\q[T]]$ induces the same winner $\q[T]$ and is therefore optimal. We select $\p[T]=\type[s^T]$. The symmetric argument applies when $\q[T]>\med$, yielding:
\begin{align*}
\p[T]=\begin{cases}
\min\set{\type[s^T],\mir{\q[T]}}& \text{if } \q[T]\leq \med,\\
\max\set{\type[s^T],\mir{\q[T]}}& \text{if } \q[T]> \med.
\end{cases}\quad&w(\p[T],\q[T])=\begin{cases}
\min\set{\q[T],\mir{\q[T]}}& \text{if } \type[s^T]<\min\set{\q[T],\mir{\q[T]}},\\
\max\set{\q[T],\mir{\q[T]}} &\text{if } \type[s^T]>\max\set{\q[T],\mir{\q[T]}},\\
\type[s^T]	& \text{else}.
\end{cases}
\end{align*}

\noindent\textbf{Optimal proposal in round $T-1$}\\
We now consider the optimal proposal and voting at round $T-1$. We organise the proof in the following way. First, we derive the winner of round~$T-1$ that would maximize the expected utility of the proposer, which, as we show, is either the proposer's own peak or the \cw. Then, we determine the optimal proposals in round~$T-1$. Finally, we show that the majority of the strategic voters would vote in favour of the proposal in Lemma~\ref{lem1}.

We now state the expected utility of the proposer~$\type[s^{T-1}]$ when proposing~$\w$ in round $T-1$. To ease the notation we denote the proposers type $\type[s^{T-1}]=\type[s]$ in this subsection:
\begin{equation*}
\text{EU}_s(\w)=
\begin{cases}
\int_0^{\w} \utility{\type[s]}{\w}\pdf[v]dv + \int_{\w}^{\mir{\w}} \utility{\type[s]}{v}\pdf[v]dv + \int_{\mir{\w}}^1 \utility{\type[s]}{\mir{\w}}\pdf[v]dv & \text{ if }\w\leq \med,\\
\int_0^{\mir{\w}} \utility{\type[s]}{\mir{\w}}\pdf[v]dv + \int_{\mir{\w}}^{\w} \utility{\type[s]}{v}\pdf[v]dv + \int_{\w}^1 \utility{\type[s]}{\w}\pdf[v]dv & \text{ if }\w> \med,
\end{cases}
\end{equation*}
which can be simplified as follows:
\begin{equation*}
\text{EU}_s(\w)=
\begin{cases}
\utility{\type[s]}{\w}\cdf[w] + \int_{\w}^{\mir{\w}} \utility{\type[s]}{v}\pdf[v]dv + \utility{\type[s]}{\mir{\w}}(1-\cdf[\mir{\w}]) & \text{ if }\w\leq \med,\\
\utility{\type[s]}{\mir{\w}}\cdf[\mir{\w}] + \int_{\mir{\w}}^{\w} \utility{\type[s]}{v}\pdf[v]dv + \utility{\type[s]}{\w}(1-\cdf[\w]) & \text{ if }\w> \med.
\end{cases}
\end{equation*}
Each of the three terms in the expected utility function corresponds to the possible winners in round $T$, depending on the proposer's type in round $T$. In the case when $\w\leq\med$ and for $\type[s^T]\in[0, \w[T-1])$, the final winner is $\w[T]=\q[T]=\w[T-1]$. For $\type[s^T]\in [\w[T-1], \mir{\w[T-1]}]$, the final winner will be \type[s^T]. Finally, for $\type[s^T]\in (\mir{\w[T-1]},1]$, the final winner is \mir{\w[T-1]}. The analogous argument holds for the case when $\w\geq\med$. 

We proceed by finding the utility-maximizing winner considering the following cases.\\
\textbf{Case 1}: $\w[T-1]\in[0, 2\med-1]$. Thus, $\mir{\w[T-1]}=1$. 
The first order condition is given by
\begin{align*}
\utility{\type[s]}{\w}'\cdf[\w]+ \utility{\type[s]}{\w}\pdf[\w]-\utility{\type[s]}{\w}\pdf[\w]=\utility{\type[s]}{\w}'\cdf[\w]=0.
\end{align*} 
Observe that $\text{EU}_{s}(w)$ is monotonically increasing for $\type[s]\geq\max\set{2\med-1,0}$. If $\type[s]<\max\set{2\med-1,0}$, we have $\utility{\type[s]}{\w}'=0$ when $\w[T-1]=\type[s]$. Thus, $\text{EU}_s(\w)$ has a maximum at $\w[T-1]=\type[s]$ as $\utility{\type[s]}{\w}'=2(\type[s]-w)$ and
\begin{equation*}
\w[T-1]=\min\set{2\med-1,\type[s]}.
\end{equation*}
\textbf{Case 2:} $\w[T-1]\in\big(\max\set{0,2\med-1},\ \min\set{2\med,1}\big)$. Let
$h:=|\w[T-1]-\med|$ and $H:=\min\set{\med,\,1-\med}$, so that $h\in[0,H]$. On this range
$\mir{\w[T-1]}=2\med-\w[T-1]$. For $h\in[0,H]$ both $\med-h$ and $\med+h$ lie in $[0,1]$ with
$\mir{\med-h}=\med+h$ and $\mir{\med+h}=\med-h$, so by the characterisation of the final round
the winner is the round-$T$ proposer's peak truncated to $[\med-h,\med+h]$ for both
$\w[T-1]=\med-h$ and $\w[T-1]=\med+h$. Both proposals thus induce the same winner lottery
$X_{h}:=\min\set{\max\set{\type[s^{T}],\,\med-h},\ \med+h}$, which equals $\type[s^{T}]$ on
$[\med-h,\med+h]$ and is pushed to the nearer endpoint outside it, while $\w[T-1]=\med$
($h=0$) yields the certain outcome $\med$. It therefore suffices to compare
$\utility{\type[s]}{\med}$ with $\text{EU}_{s}(\w[T-1])=-\mathbb{E}[(\type[s]-X_{h})^{2}]$:
\[
\text{EU}_{s}(\med)-\text{EU}_{s}(\w[T-1])
=\mathbb{E}\big[(X_{h}-\med)^{2}\big] - 2(\type[s]-\med)\,\mathbb{E}\big[X_{h}-\med\big].
\]
Since $X_{h}=\min\set{\max\set{\type[s^{T}],\med-h},\med+h}$, its centred value
$X_{h}-\med$ equals $-h$ when $\type[s^{T}]<\med-h$, equals $\type[s^{T}]-\med$ when
$\type[s^{T}]\in[\med-h,\med+h]$, and equals $+h$ when $\type[s^{T}]>\med+h$. Hence
\begin{align*}
\mathbb{E}\big[X_{h}-\med\big]
&=-h\,\cdf[\med-h]
+\int_{\med-h}^{\med+h}(v-\med)\,\pdf[v]\,dv
+h\,\big(1-\cdf[\med+h]\big),\\
\mathbb{E}\big[(X_{h}-\med)^{2}\big]
&=h^{2}\,\cdf[\med-h]
+\int_{\med-h}^{\med+h}(v-\med)^{2}\,\pdf[v]\,dv
+h^{2}\,\big(1-\cdf[\med+h]\big).
\end{align*}
Integrating the middle term of the first line by parts,
$\int_{\med-h}^{\med+h}(v-\med)\pdf[v]\,dv
=\big[(v-\med)\cdf[v]\big]_{\med-h}^{\med+h}-\int_{\med-h}^{\med+h}\cdf[v]\,dv
=h\,\cdf[\med+h]+h\,\cdf[\med-h]-\int_{\med-h}^{\med+h}\cdf[v]\,dv$,
and the boundary terms cancel those already present, leaving
\[
\mathbb{E}\big[X_{h}-\med\big]=h-\int_{\med-h}^{\med+h}\cdf[v]\,dv .
\]
Substituting $v=\med+r$ and folding the two half-ranges, $\int_{\med-h}^{\med+h}\cdf[v]\,dv
=\int_{0}^{h}\big(\cdf[\med+r]+\cdf[\med-r]\big)\,dr$, while $h=\int_{0}^{h}dr$; therefore
\[
\mathbb{E}\big[X_{h}-\med\big]
=\int_{0}^{h}\big(1-\cdf[\med+r]-\cdf[\med-r]\big)\,dr
=\int_{0}^{h} q(r)\,dr .
\]
Treating the second line identically---by parts,
$\int_{\med-h}^{\med+h}(v-\med)^{2}\pdf[v]\,dv
=\big[(v-\med)^{2}\cdf[v]\big]_{\med-h}^{\med+h}-2\int_{\med-h}^{\med+h}(v-\med)\cdf[v]\,dv$,
whose boundary terms again cancel---gives
\[
\mathbb{E}\big[(X_{h}-\med)^{2}\big]
=h^{2}-2\int_{\med-h}^{\med+h}(v-\med)\,\cdf[v]\,dv
=2\int_{0}^{h} r\,\big(1-\cdf[\med+r]+\cdf[\med-r]\big)\,dr
=2\int_{0}^{h} r\,s(r)\,dr,
\]
where the last line substitutes $v=\med+r$, uses $h^{2}=2\int_{0}^{h}r\,dr$, and folds the two
half-ranges of $\int_{\med-h}^{\med+h}(v-\med)\cdf[v]\,dv=\int_{0}^{h}r\big(\cdf[\med+r]-\cdf[\med-r]\big)\,dr$.
Substituting these moments gives
\begin{equation*}
\text{EU}_{s}(\med)-\text{EU}_{s}(\w[T-1])
=2\left(\int_{0}^{h} r\,s(r)\,dr-(\type[s]-\med)\int_{0}^{h} q(r)\,dr\right),
\qquad \type[s]\in[0,1],\ h\in[0,H].
\end{equation*}
 
Since $\type[s]\in[\lbound{\type},\ubound{\type}]$, we have
$\type[s]-\med\in[-(\med-\lbound{\type}),\,\ubound{\type}-\med]$, and therefore
\[
-(\type[s]-\med)\int_{0}^{h} q(r)\,dr\ \geq\
\begin{cases}
-(\ubound{\type}-\med)\displaystyle\int_{0}^{h} q(r)\,dr, & \text{if } \displaystyle\int_{0}^{h} q(r)\,dr\geq 0,\\[10pt]
\phantom{-}(\med-\lbound{\type})\displaystyle\int_{0}^{h} q(r)\,dr, & \text{if } \displaystyle\int_{0}^{h} q(r)\,dr\leq 0.
\end{cases}
\]
In either case, Assumption~\ref{assumption:F3} gives
$\int_{0}^{h} r\,s(r)\,dr-(\type[s]-\med)\int_{0}^{h} q(r)\,dr\geq 0$ for every $h\in[0,H]$,
hence $\text{EU}_{s}(\med)-\text{EU}_{s}(\w[T-1])\geq 0$. Therefore, for every $\type[s]\in[\lbound{\type},\ubound{\type}]$, the median peak yields weakly higher expected utility than every winner in the middle range.

\textbf{Case 3}: $\w[T-1]\in [2\med,1]$. Thus, by definition $\mir{\w[T-1]}=0$.
The FOC is given by
\begin{align*}
\begin{gathered}
\utility{\type[s]}{\w}'(1-\cdf[\w])+ \utility{\type[s]}{\w}\pdf[\w]-\utility{\type[s]}{\w}\pdf[\w]=\utility{\type[s]}{\w}'(1-\cdf[\w])=0.
\end{gathered}
\end{align*} 
Observe that $\text{EU}_s(w)$ is monotonically decreasing for $\type[s]\leq\min\set{2\med,1}$. If $\type[s]>\min\set{2\med,1}$, the expected utility is maximized at $\w[T-1]=\type[s]$. Thus, 
\begin{equation*}
\w[T-1]=\max\set{2\med,\type[s]}.
\end{equation*}
The utility-maximizing winner of round~$T-1$ therefore lies in
$\set{\min\set{2\med-1,\type[s]},\ \med,\ \max\set{2\med,\type[s]}}$.
We organise the rest of the proof in a series of lemmas:
\begin{enumerate}
        \item we identify the interval of types of proposers who have a higher expected utility from $\type[s]$ winning round~$T-1$ instead of $\med$ when $\type[s]< 2\med-1$ or $\type[s]> 2\med$,
        \item we show that all such types of proposers prefer a winner within the interval over $\med$,
        \item we determine the voting behaviour that sustains the resulting proposal, and show that a decisive majority votes identically whether sincerely or sophisticatedly.
    \end{enumerate}

\begin{lemma}
If $\type[s]\in[0,\lbound{\type})\cup(\ubound{\type},1]$, then $\text{EU}_{s}(\type[s])>\utility{\type[s]}{\med}$. If $\type[s]\in[\lbound{\type},\ubound{\type}]$, then $\text{EU}_{s}(\type[s])\leq\utility{\type[s]}{\med}$.
\label{lem1}
\end{lemma}

\begin{proof}
Suppose first that $\med>\tfrac12$ and define
\[
D_L(\type[s])
:=\text{EU}_{s}(\type[s])-\utility{\type[s]}{\med}
=\int_{\type[s]}^1\utility{\type[s]}{v}\pdf[v]\,dv-
\utility{\type[s]}{\med}
\]
for $\type[s]\leq2\med-1$. Differentiating gives
\[
D_L'(\type[s])
=2\left(\type[s]F(\type[s])+
\int_{\type[s]}^1v\pdf[v]\,dv-\med\right).
\]
At $\type[s]=0$ the expression in parentheses is nonpositive since
$\mean\leq\med$ whenever $\med>\tfrac12$, by
Assumption~\ref{assumption:F} and Lemma~\ref{lem-mean-median}. It is
nondecreasing in $\type[s]$, and at $2\med-1$ it equals
\[
\int_0^{1-\med}
\bigl(1-F(\med+h)-F(\med-h)\bigr)\,dh\leq0
\]
by Assumption~\ref{assumption:F}. Hence $D_L$ is nonincreasing on
$[0,2\med-1]$. By the definition of $\lbound{\type}$ as the boundary of
the strict-preference region, $D_L(\type[s])>0$ for
$\type[s]\in[0,\lbound{\type})$ and $D_L(\type[s])\leq0$ for
$\type[s]\in[\lbound{\type},2\med-1]$.

For $\type[s]\in[2\med-1,1]$, the proposer's own peak belongs to the
middle range, because $2\med>1$. Case~2 and
Assumption~\ref{assumption:F3} therefore give
$\text{EU}_{s}(\type[s])\leq\utility{\type[s]}{\med}$ for every
interior type. This proves the claim when $\med>\tfrac12$.

Suppose next that $\med<\tfrac12$. The symmetric upper-tail difference
\[
D_R(\type[s])
:=\text{EU}_{s}(\type[s])-\utility{\type[s]}{\med}
=\int_0^{\type[s]}\utility{\type[s]}{v}\pdf[v]\,dv-
\utility{\type[s]}{\med}
\]
is nondecreasing on $[2\med,1]$, by the symmetric calculation and
Assumption~\ref{assumption:F}. Hence $D_R(\type[s])>0$ for
$\type[s]\in(\ubound{\type},1]$ and $D_R(\type[s])\leq0$ for
$\type[s]\in[2\med,\ubound{\type}]$. For
$\type[s]\in[0,2\med]$, the comparison follows from Case~2 and
Assumption~\ref{assumption:F3}. At $\med=\tfrac12$, Case~2 and continuity
cover the entire policy interval. This completes the proof.
\end{proof}
Next, we show that sufficiently extreme proposer types prefer every winner between their peak and the corresponding threshold to the median peak.
\begin{lemma}
For every $\type[s]\in[0,\lbound{\type})$, $\text{EU}_{s}(w)\geq\utility{\type[s]}{\med}$ for all $w\in[\type[s],\lbound{\type}]$,
and, symmetrically, for every $\type[s]\in(\ubound{\type},1]$, $\text{EU}_{s}(w)\geq\utility{\type[s]}{\med}$ for all $w\in[\ubound{\type},\type[s]]$.
\label{lem2}
\end{lemma}

\begin{proof}
Suppose first that $\med>\tfrac12$. If $\lbound{\type}=0$, the lower-tail
claim is vacuous. Let $\lbound{\type}>0$ and fix
$\type[s]\in[0,\lbound{\type})$. Case~1 implies that
$\text{EU}_{s}(w)$ is nonincreasing in $w$ on
$[\type[s],\lbound{\type}]$, so it suffices to compare
$\lbound{\type}$ with $\med$. Define
\[
K_L(\type[s])
:=\text{EU}_{\type[s]}(\lbound{\type})-
\utility{\type[s]}{\med}.
\]
Differentiating with respect to $\type[s]$ gives
\[
K_L'(\type[s])
=2\left(
\lbound{\type}\cdf[\lbound{\type}]
+\int_{\lbound{\type}}^1v\pdf[v]\,dv-
\med
\right).
\]
The expression in parentheses is nonpositive. Indeed, it is
nondecreasing in $\lbound{\type}\leq2\med-1$, and its value at
$2\med-1$ is nonpositive by the calculation in the proof of
Lemma~\ref{lem1}. Since $K_L(\lbound{\type})=0$, it follows that
$K_L(\type[s])\geq0$ for every
$\type[s]\in[0,\lbound{\type})$. Together with the monotonicity in $w$,
this proves the lower-tail claim.

The upper-tail claim when $\med<\tfrac12$ is symmetric. If
$\ubound{\type}=1$, it is vacuous. Otherwise, Case~3 shows that
$\text{EU}_{s}(w)$ is nondecreasing on
$[\ubound{\type},\type[s]]$. The symmetric analogue of $K_L$ is
nondecreasing in $\type[s]$ and vanishes at $\ubound{\type}$, so it is
nonnegative for every $\type[s]\in(\ubound{\type},1]$.
\end{proof}

It remains to translate the utility-maximizing winner into a proposal. If $\q[T-1]\geq\lbound{\type}$, no proposal below $\lbound{\type}$ can win against the standing status quo. Thus a lower-tail winner is attainable only if both the status quo and the proposer's type are below $\lbound{\type}$. Symmetrically, an upper-tail winner is attainable only if both are above $\ubound{\type}$. In every other case, the best attainable winner is $\med$. Hence an optimal proposal for the proposer in round $T-1$ is:
\begin{equation}
\p[T-1]=
\begin{cases}
\type[s^{T-1}]& \text{ if }\max\set{\q[T-1], \type[s^{T-1}]}< \lbound{\type} \text{ or } \min\set{\q[T-1],\type[s^{T-1}]}>\ubound{\type},\\
\med & \text{else.}
\end{cases}
\label{eq-prop}
\end{equation}

It remains to verify that this proposal is sustained by the voting behaviour selected by the pivotal-voting refinement, and that the same winner is obtained under sincere voting.

\medskip
\noindent\emph{Interior-type bound.}
For every $\type\in[\lbound{\type},\ubound{\type}]$ and every $w\in[0,1]$,
\[
\text{EU}_{\type}(w)\leq\utility{\type}{\med}.
\]
To see this, first suppose that $\med>\tfrac12$. For a winner in the middle range, the claim follows from Case~2 and Assumption~\ref{assumption:F3}. For $w\leq2\med-1$, Case~1 shows that the maximum over this range is attained at $w=\type$ when $\type\leq2\med-1$, and at $w=2\med-1$ otherwise. The first comparison follows from Lemma~\ref{lem1}; the second follows from Case~2 at the boundary, by continuity. Since $2\med>1$, there is no upper outer range. The case $\med<\tfrac12$ is symmetric, and when $\med=\tfrac12$ Case~2 and continuity cover the whole interval.

\begin{lemma}
\label{lem3}
Under the proposal strategy in Equation~\eqref{eq-prop}, sophisticated voting satisfies the pivotal-voting refinement. Under the same proposal strategy, sincere and sophisticated voting induce the same winner in every round.
\end{lemma}
\begin{proof}
Suppose first that $\p[T-1]=\med$. If $\med>\tfrac12$, the set $[\lbound{\type},1]$ has mass $1-\cdf[\lbound{\type}]>\tfrac12$. By the interior-type bound, every voter in this set weakly prefers the certain continuation outcome $\med$ to the round-$T$ lottery induced by $\q[T-1]$. Hence $\med$ wins under sophisticated voting. Under sincere voting, $\med$ also wins because it is the Condorcet winner. Thus the winner is the same under the two voting behaviours, although their supporting coalitions need not coincide. The case $\med<\tfrac12$ is symmetric, using the decisive set $[0,\ubound{\type}]$; at $\med=\tfrac12$ the threshold interval is the whole policy space.

Next suppose that $\p[T-1]=\type[s^{T-1}]$. If $\med>\tfrac12$, both $\q[T-1]$ and $\type[s^{T-1}]$ lie below $\lbound{\type}$. Let
\[
a:=\max\set{\q[T-1],\type[s^{T-1}]},
\qquad
b:=\min\set{\q[T-1],\type[s^{T-1}]}.
\]
Every voter with type at least $a$ weakly prefers the round-$T$ lottery induced by $a$ to the lottery induced by $b$: for every final-round proposer type, the former continuation outcome is weakly higher, and the two outcomes differ only when the higher one is closer to such a voter. This set has mass $1-\cdf[a]>\tfrac12$, so $a$ wins under sophisticated voting. The same set sincerely supports $a$ against $b$, and therefore the same alternative wins under sincere voting. The argument for $\med<\tfrac12$ is symmetric, with the lower of the two alternatives supported by the decisive set of types below it. The same reasoning applies recursively in every earlier round because the proposal rule preserves the relevant tail interval until $\med$ is proposed.
\end{proof}

\noindent\textbf{Optimal proposal in rounds $t \in \{1,2,\dots,T-2\}$}\\
If $\lbound{\type}=0$ and $\ubound{\type}=1$, the preceding analysis implies that every proposer selects the \cw. Suppose instead that $\med>\tfrac12$ and $\lbound{\type}>0$; the case $\med<\tfrac12$ is symmetric. If the standing status quo is at least $\lbound{\type}$, no lower-tail proposal can win, and the continuation outcome is $\med$. If the status quo is below $\lbound{\type}$, a proposer compares a lower-tail continuation with the certain outcome $\med$.

For a proposer of type $\type[s^t]$, the expected-utility difference generated by the boundary winner $\lbound{\type}$ in round $t\leq T-2$ is
\begin{align*}
&\text{EU}_{\type[s^t]}(\lbound{\type})-\utility{\type[s^t]}{\med}\\
&\quad=
\utility{\type[s^t]}{\lbound{\type}}\cdf[\lbound{\type}]^{T-t}
+\cdf[\lbound{\type}]^{T-t-1}
\int_{\lbound{\type}}^1\utility{\type[s^t]}{v}\pdf[v] \,dv
-\utility{\type[s^t]}{\med}\cdf[\lbound{\type}]^{T-t-1}\\
&\quad=
\cdf[\lbound{\type}]^{T-t-1}
\left(
\utility{\type[s^t]}{\lbound{\type}}\cdf[\lbound{\type}]
+\int_{\lbound{\type}}^1\utility{\type[s^t]}{v}\pdf[v] \,dv
-\utility{\type[s^t]}{\med}
\right).
\end{align*}
The term in parentheses is the round-$(T-1)$ comparison at $\lbound{\type}$. It is nonnegative for $\type[s^t]\in[0,\lbound{\type})$ by Lemma~\ref{lem2}. It is nonpositive for $\type[s^t]\in[\lbound{\type},\med]$ because, as established in the proof of Lemma~\ref{lem2}, it is nonincreasing in the proposer's type and vanishes at $\lbound{\type}$. Together with the monotonicity of the lower-tail continuation in the winner, this gives the same proposal rule as in Equation~\eqref{eq-prop}. The upper-tail argument is symmetric.

Hence the rule in Equation~\eqref{eq-prop} selects an optimal proposal in every round $t\leq T-1$. Finally, the argument of Lemma~\ref{lem3} applies recursively in every earlier round: sincere and sophisticated voting induce the same winner at each voting stage.

\begin{lemma}
\label{lem-endpoint}
Let Assumptions~\ref{assumption:F} and~\ref{assumption:F3} hold. Then
$\text{EU}_{2\med-1}(2\med-1)
\leq
\utility{2\med-1}{\med}$ if $\med>\tfrac12$, and
$\text{EU}_{2\med}(2\med)
\leq
\utility{2\med}{\med}$ if $\med<\tfrac12$.
\end{lemma}
\begin{proof}
Write
\[
Q_H:=\int_0^Hq(r)\,dr,
\qquad
R_H:=\int_0^Hr\,s(r)\,dr.
\]
Suppose first that $\med>\tfrac12$ and let $H:=1-\med$. The moment calculation in Case~2, evaluated at $\type[s]=w=2\med-1$, gives $\text{EU}_{2\med-1}(2\med-1)-\utility{2\med-1}{\med}
=-2\bigl(R_H+HQ_H\bigr)$. Assumption~\ref{assumption:F} gives $Q_H\leq0$. Moreover, $\lbound{\type}\leq2\med-1$, so $\med-\lbound{\type}\geq H$. By Assumption~\ref{assumption:F3}, $R_H\geq-(\med-\lbound{\type})Q_H\geq-HQ_H$, which proves the first inequality.

Suppose next that $\med<\tfrac12$ and let $H:=\med$. At $\type[s]=w=2\med$, the same moment calculation gives$\text{EU}_{2\med}(2\med)-\utility{2\med}{\med}
=2\bigl(HQ_H-R_H\bigr)$.
Now Assumption~\ref{assumption:F} gives $Q_H\geq0$, while $\ubound{\type}-\med\geq H$. Assumption~\ref{assumption:F3} therefore implies $R_H\geq(\ubound{\type}-\med)Q_H\geq HQ_H$, which proves the second inequality.
\end{proof}

\section{Appendix: Proof of Theorem~\ref{prop-2rounds}}
\label{proof-2rounds}
We prove the two parts of the theorem in turn.

\noindent\textbf{Part (i).}
Let $\q[1]\notin[\lbound{\type},\ubound{\type}]$. Consider $\med>\tfrac12$, so that $\ubound{\type}=1$ and hence $\q[1]<\lbound{\type}$. Whenever the \cw is proposed in a round $t\leq T-1$, it wins that round and is implemented. The only history in which $\med$ is not proposed in the first $T-1$ rounds is therefore the history in which every proposer in those rounds has a type in $[0,\lbound{\type})$. Along that history the status quo remains in the lower-tail interval. Since proposer types are drawn independently from $\cdf[]$, this event has probability $\cdf[\lbound{\type}]^{T-1}$, and hence
\[
\operatorname{Prob}(\w[T]=\med)
=1-\cdf[\lbound{\type}]^{T-1}.
\]
The symmetric argument for $\med<\tfrac12$ yields
$1-(1-\cdf[\ubound{\type}])^{T-1}$. If
$\q[1]\in[\lbound{\type},\ubound{\type}]$, the first proposer selects $\med$, which is implemented. When $\med=\tfrac12$, the threshold interval is $[0,1]$, so only this latter case arises.

\noindent\textbf{Part (ii).}
We first show that Assumption~\ref{assumption:F2} is equivalent to $\lbound{\type}=0$ when $\med>\tfrac12$, and to $\ubound{\type}=1$ when $\med<\tfrac12$. In either case the other threshold is already at the boundary, so $[\lbound{\type},\ubound{\type}]=[0,1]$. Equation~\eqref{eq-prop} then prescribes $\med$ in round~1, and Lemma~\ref{lem3} implies that it wins under sophisticated and sincere voting alike. Hence $\q[2]=\med$ and $\med$ is implemented when $T=2$. At $\med=\tfrac12$, the threshold interval is $[0,1]$ by definition. Moreover, Assumption~\ref{assumption:F} gives $\mean\geq\med$, so $\med^2-\mean^2\leq0$ and the first branch of Assumption~\ref{assumption:F2} holds automatically.

\medskip
\noindent\emph{Case $\med>\tfrac12$.}
Let
\[
D_L(\type[s])
:=\text{EU}_{\type[s]}(\type[s])-\utility{\type[s]}{\med}.
\]
As established in the proof of Lemma~\ref{lem1}, $D_L$ is continuous and nonincreasing on $[0,2\med-1]$. Lemma~\ref{lem-endpoint} gives $D_L(2\med-1)\leq0$. By the definition of $\lbound{\type}$ as the boundary of the strict-preference region,
\[
\lbound{\type}=0
\quad\Longleftrightarrow\quad
D_L(0)\leq0.
\]
Since $\mir{0}=1$ and $\utility{0}{0}=0$,
\[
D_L(0)
=\text{EU}_{0}(0)-\utility{0}{\med}
=\int_0^1\utility{0}{v}\pdf[v] \,dv-\utility{0}{\med}.
\]
Thus $\lbound{\type}=0$ if and only if
$\int_0^1v^2\pdf[v] \,dv\geq\med^2$, equivalently, $\operatorname{Var}(\type)\geq\med^2-\mean^2$, which is Assumption~\ref{assumption:F2} in this case.

\medskip
\noindent\emph{Case $\med<\tfrac12$.}
Define
\[
D_R(\type[s])
:=\text{EU}_{\type[s]}(\type[s])-\utility{\type[s]}{\med}.
\]
The symmetric calculation in the proof of Lemma~\ref{lem1} shows that $D_R$ is continuous and nondecreasing on $[2\med,1]$, while Lemma~\ref{lem-endpoint} gives $D_R(2\med)\leq0$. Hence
\[
\ubound{\type}=1
\quad\Longleftrightarrow\quad
D_R(1)\leq0
\quad\Longleftrightarrow\quad
\int_0^1\utility{1}{v}\pdf[v] \,dv\leq\utility{1}{\med}.
\]
Substituting the utility function and simplifying gives
\begin{align*}
\operatorname{Var}(\type)+\mean^2-\med^2
&\geq2(\mean-\med),\\
\operatorname{Var}(\type)
&\geq(\mean-\med)(2-\mean-\med)\\
&=(1-\med)^2-(1-\mean)^2,
\end{align*}
which is Assumption~\ref{assumption:F2} for $\med<\tfrac12$.

It remains to establish the realisation-level sufficient condition. Suppose that
$\q[1]\in[\lbound{\type},\ubound{\type}]$ or
$\type[s^1]\in[\lbound{\type},\ubound{\type}]$. Equation~\eqref{eq-prop} prescribes the proposer's own type only if
\[
\max\set{\q[1],\type[s^1]}<\lbound{\type}
\quad\text{or}\quad
\min\set{\q[1],\type[s^1]}>\ubound{\type}.
\]
Either membership condition makes both inequalities false, so $\p[1]=\med$. By Lemma~\ref{lem3}, $\med$ wins round~1 under sophisticated and sincere voting alike, and therefore $\q[2]=\med$. Since $\mir{\med}=\med$, the final-round winner is $\med$. Hence $\w[2]=\med$ for every such realisation.
%
%
\end{document}